\documentclass[structabstract,a4paper]{aa}  
\usepackage{graphicx,amsmath}
\usepackage{txfonts}
\begin{document}
\newcommand{\mearth}{M_\oplus}
\newcommand{\tacc}{t_{acc}}
\newcommand{\tdyn}{t_{dyn}}
\newcommand{\hdz}{H_{DZ}}
\title{On the growth and orbital evolution of giant planets in layered 
protoplanetary disks}

   \author{A. Pierens \inst{1,2}
          \and
          R.P. Nelson \inst{3}
          }

   \offprints{A. Pierens}

   \institute{ Universit\'e de Bordeaux, Observatoire Aquitain des Sciences de l'Univers, 
    BP89 33271 Floirac Cedex, France \label{inst1} \and
   Laboratoire d'Astrophysique de Bordeaux,
    BP89 33271 Floirac Cedex, France \label{inst2} \\
     \email{arnaud.pierens@obs.u-bordeaux1.fr}       
  \and Astronomy Unit, Queen Mary, University of London, Mile End Road, London, E1 4NS, UK 
   \label{inst3} }

   \date{}

 
  \abstract
   {}
   {We present the results of hydrodynamic simulations of the growth
    and orbital
    evolution of giant planets embedded in a protoplanetary disk 
    with a dead-zone. The aim is 
    to examine to what extent the presence of a dead-zone affects 
    the rates of mass accretion and migration for giant planets.}
   {We performed 3D numerical simulations using a grid--based hydrodynamics code. 
    In these simulations of laminar, non-magnetised disks, 
    the dead-zone is treated as a region where the vertical profile of
    the viscosity depends on the distance from the equatorial plane.
    We consider dead-zones with vertical sizes, $\hdz$, ranging from $0$ to $\hdz=2.3H$,
    where $H$ is the disk scale-height. For all models, the vertically
    integrated viscous stress, and the related mass flux through 
    the disk, have the same value (equivalent to $10^{-8}$ M$_{\odot}$yr$^{-1}$),
    such that the simulations test the dependence of planetary mass accretion and
    migration on the vertical distribution of the viscous stress (and mass flux).
    For each model, an embedded $30\;M_\oplus$ planet on a fixed circular orbit
    is allowed to accrete gas from the disk.
    Once the planet mass becomes equal to that of Saturn or Jupiter,
    we allow the planet orbit to evolve due to gravitational interaction with
    the disk.}
   {We find that the time scale over which a protoplanet grows to become 
    a giant planet is essentially independent of the dead-zone size, 
    and depends only on the total rate at which the disk viscously
    supplies material to the planet. For Saturn-mass planets, the 
    migration rate depends only weakly on the size of the dead-zone
    for $\hdz\le 1.5H$, but becomes noticeably slower when $\hdz=2.3H$.
    This effect is apparently due to the desaturation of corotation
    torques which originate from residual material in the partial-gap region.
    For Jupiter-mass planets, there is a clear tendency for the migration
    to proceed more slowly as the size of the dead-zone increases,
    with migration rates differing by approximately 40 \% for models
    with $\hdz=0$ and $\hdz=2.3H$.}
   {Our results indicate that for disks models in which the mass accretion rate
    has a well defined value, the accretion and migration rates 
    for Saturn- and Jovian-mass planets are relatively insensitive to
    the presence and size of a dead-zone.}

   \keywords{accretion, accretion disks --
                planetary systems: formation --
                hydrodynamics --
                methods: numerical
               }

   \maketitle
%

\section{Introduction}
Multiple observations of star-forming regions have revealed that young stars are surrounded by 
circumstellar disks composed
of gas and dust (e.g. Beckwith  1996, Sicilia-Aguilar et al. 2006), 
which are thought to provide the necessary material for planet formation. These 
protoplanetary disks generally show evidence for gas accretion onto the central star, with a
typical mass flow being $\sim 10^{-8}$ M$_{\oplus}$ yr$^{-1}$ 
(Sicilia-Aguilar et al. 2004; Gullbring
et al. 1998; Haisch et al. 2001). It has long been recognized that 
such a value for the 
accretion rate requires a source of anomalous viscosity, 
probably originating from turbulence, 
to efficiently transport the angular momentum outward (Shakura \& Sunyaev 1973). \\
Although instabilities such as the baroclinic instability 
(Klahr \& Bodenheimer 2003) or the 
Rossby-wave instability (Lovelace et al. 1999) may result in turbulence, 
a significant number of studies (Hawley et al. 1996; Brandenburg et al. 1996)
have shown that the non-linear outcome of
the  magnetorotational instability (Balbus \& Hawley 1991; hereafter MRI) 
is MHD turbulence 
with an effective viscous stress parameter $\alpha \sim 5\times10^{-3}$, 
providing 
the necessary angular momentum transport to match the 
observed accretion rates onto T-Tauri 
stars (Hartmann et al. 1998). \\
Although the MRI has been proven robust in generating MHD turbulence, the 
fact that the 
ionisation fraction is expected to be low in 
protoplanetary disks (Blaes \& Balbus 1994) raises 
a number of question about the applicability of 
the MRI in these environments. That is the 
reason why it has been suggested (Gammie 1996) that protoplanetary disks 
may have both 
magnetically active zones near the surface, where cosmic ray 
ionisation enables the MRI to 
develop, and a ``dead-zone'' close to the disk midplane where the 
flow remains laminar. This 
layered-disk picture was confirmed by Fleming \& Stone (2003) 
who performed MHD simulations of 
stratified disks in which the magnetic resistivity decreases 
as a function of height. Over the 
past few years, many studies have focused on the structure of 
the dead-zone by using different 
chemical reaction networks. Fromang et al. (2002) showed that 
the size of the dead-zone can 
decrease in the presence of heavy metals (such as magnesium) due to charge-transfer 
reactions. Ilgner \& Nelson (2006b) 
examined the role of turbulent mixing, and showed that in the 
presence of magnesium, such a process 
can enliven the dead-zone beyond a few AU. This was confirmed by 
Ilgner \& Nelson (2008) who 
performed multifluid MHD simulations and who showed that 
despite the addition of gas-phase 
magnesium,  dead-zones in protoplanetary disks typically persist 
at distances $R\le 5$  AU, 
with potentially important consequences for planet formation 
in these regions. Calculations which examine the sizes of dead-zones
including gas-grain chemistry, show that the presence of
a modest population of sub-micron sized grains cause the planet
forming regions of protoplanetary disks to host significant dead-zones
(Sano et al. 2000; Ilgner \& Nelson 2006a; Turner \& Drake 2009).    \\
So far, the effect of a dead-zone on planet formation has received 
little attention, in large part because of the computational expense
of performing large-scale 3D MHD simulations of turbulent protoplanetary
disks. A plausible
scenario for the formation of planets involves the following steps: 
i) coagulation and settling of dust in the disk midplane, followed by
the growth of km-sized planetesimals;
ii) runaway growth of planetesimals (e.g. Greenberg et al. 1978; 
Wetherill \& Stewart 1989) into $\sim 10^{-5}\;M_\oplus$ embryos;
iii) oligarchic growth of these embryos (e.g. 
Kokubo \& Ida 1998, 2000; Leinhardt \& Richardson 2005) 
into planetary cores. Planetary cores
forming oligarchically beyond the snow-line are expected 
to have masses $\sim 10\; M_\oplus$ 
(Thommes et al. 2003) and consequently are able to accrete 
a gaseous envelope to become giant 
planets (Pollack et al. 1996). 

The impact of a dead-zone 
on dust settling was examined by 
Fromang \& Papaloizou (2006) who found  a tendency for 
thinner dust sub-disks to form in the 
presence of a dead-zone. Lyra \& al. (2008, 2009) found that a 
Rossby-wave instability can be 
triggered at the border of a dead-zone where the surface density 
is significantly enhanced, 
creating vortices which are efficient in trapping solids and 
forming planetary embryos. \\
The consequence of a dead-zone on planet migration was 
studied by Matsumura et al. (2003). 
These authors found that a dead-zone significantly slows down the 
migration of low-mass planets undergoing Type I migration.
This arises because the dead-zone
creates a jump in surface density, which changed the
balance between inner and outer Lindblad torques in their
model. A surface density transition, however, can also
create a planet trap where the corotation torque exerted on the
protoplanet equals the differential Lindblad torque 
(Masset et al. 2006). Concerning 
gap-opening giant planets which migrate on a viscous timescale, 
Matsumura et al. (2003) found, unsurprisingly, that 
a dead-zone slows down  Type II migration 
due to the low value for the viscosity there. However, it should 
be noted that this work did not consider how the existence of a 
live-zone near the disk surface affects this latter result. \\
In this paper, we present the results of 3D hydrodynamical simulations of 
giant-planets 
embedded in layered protoplanetary disks and 
in which the dead-zone is modelled using a 
vertical profile for the laminar viscosity which 
increases as a function of height in the disk. 
The aim of this work is to investigate how the evolution depends on 
the size of the dead-zone 
$\hdz$, subject to the condition that all our disk models
have the same vertically integrated viscous stress and 
associated radial mass flux.  To address this issue, we consider a model in 
which a $30\;M_\oplus$  
protoplanet is embedded in the disk on a fixed circular orbit,
and can slowly accrete gas until its mass $m_p$ 
becomes the same as that of Saturn ($m_p=1$ M$_S$) or Jupiter ($m_p=1$ M$_J$). 
The planet is then released and allowed to evolve under the action of 
disk forces. The results of these 
simulations suggest that for a given accretion rate through 
the disk, Jupiter migrates more 
slowly as the size of the dead-zone increases. However, we find 
that provided the size of the 
dead-zone is small enough, both the accretion rate onto 
the planet and the migration of 
Saturn-mass planets depend only weakly on $\hdz$. \\
	    
This paper is organized as follows. In Sect. 2, we describe the 
hydrodynamical model and the 
numerical setup. In Sect. 3, we present the results of our simulations. 
We finally summarize 
and draw our conclusions in Sect. 4.


\section{Physical model and numerical setup}

In order to study the evolution of a giant planet in a dead-zone, 
which is represented in the 
simulations as a disk region where the viscosity profile depends 
on the distance from the 
equatorial plane, we adopt a 3-dimensional disk model. In spherical coordinates 
$(r,\theta,\phi)$ and in a frame centred on the central star, the continuity equation reads:

\begin{equation}
\frac{\partial \rho}{\partial t}+\nabla\cdot(\rho {\bf v})=0
\end{equation}

\noindent
where $\rho$ is the disk density. The equations for the radial, 
meridional, and angular 
components of the disk velocity ${\bf v}=(v_r,v_\theta,v_\phi)$ are given,
respectively, by:

\begin{equation}
\frac{\partial (\rho v_r)}{\partial t}+\nabla \cdot (\rho v_r {\bf v})=
\frac{\rho(v_\theta^2+v_\phi^2)}{r}-\frac{\partial p}{\partial r}-
\rho\frac{\partial \Phi}{\partial r}+f_r,
\end{equation}

\begin{equation}
\frac{\partial (\rho v_\theta)}{\partial t}+\nabla \cdot (\rho v_\theta 
{\bf v})= -\frac{\rho v_rv_\theta}{r}-\frac{\rho v_\phi^2 \cot \theta}
{r}-\frac{1}{r}\frac{\partial p}{\partial r}-\frac{\rho}{r}\frac{\partial 
\Phi}{\partial r}+f_\theta
\end{equation}

\noindent
and

\begin{align}
\frac{\partial (\rho v_\phi)}{\partial t}+\nabla \cdot (\rho v_\phi 
{\bf v})=& -\frac{\rho v_rv_\phi}{r}-\frac{\rho v_\theta v_\phi \cot \theta}
{r}\\
&-\frac{1}{r\sin \theta}\frac{\partial p}{\partial r}-\frac{\rho}{r\sin 
\theta}\frac{\partial \Phi}{\partial r}+f_\phi.
\end{align}

\noindent
In the above equations, $p$ is the pressure, 
$f_r$, $f_\theta$ and $f_\phi$ are respectively 
the radial, meridional and azimuthal components of the viscous force per unit volume. 
Expressions for $f_r$, $f_\theta$ and $f_\phi$ can be found 
for example in Klahr et al.(1999). 
$\Phi$ is the gravitational potential and can be written as:
\begin{equation}
\Phi=-\frac{GM_\star}{r}-\frac{Gm_p}{\sqrt{|{\bf r}-{\bf r_p}|^2+\epsilon ^2}}
+\Phi_{ind},
\end{equation}
where $M_\star$ and $m_p$ are the masses of the star and 
planet, respectively, and where 
$\epsilon$ is a softening length. $\Phi_{ind}$ is an 
indirect term arising from the fact that 
the star-centered frame is not inertial. This term reads:
\begin{equation}
\Phi_{ind}=\frac{Gm_p}{r_p^3}{\bf r \cdot r_p}+G\int_V\frac{dm({\bf r'})
}{r'^3}{\bf r \cdot r'},
\end{equation}
where the integral is performed over the volume of the disk.

\subsection{Planet orbital evolution}

In this work the planet can experience the 
gravitational acceleration arising from both the 
disk and the central star. Therefore, the 
equation of motion for the planet is given by:
\begin{equation}
\frac{d^2{\bf r_p}}{dt^2}=-\frac{GM_\star}{r_p^3}{\bf r_p}+{\bf f_{dp}}-
\nabla \Phi_{ind},
\end{equation}
where ${\bf f_{dp}}$ is the force due to the disk which is defined by:
\begin{equation}
{\bf f_{dp}}=-G\int_V\frac{\rho(\bf{r})({\bf r_p}-{\bf r})}
{(|{\bf r}-{\bf r_p}|^2+\epsilon^2)^{3/2}}dV
\end{equation}
Note that we exclude the material contained in the planet Hill 
sphere when calculating 
the gravitational force acting on the planet. Moreover, in the simulations 
presented here, the smoothing 
length $\epsilon$ is set to the diagonal length of one grid cell.

\section{Numerical setup}
\subsection{Numerical method}

The 3D hydrodynamical simulations presented in this paper have
been performed using the NIRVANA 
code (Ziegler \& Yorke 1997), which is basically 
a grid-based code which computes spatial 
derivatives using finite differences. The 
advection scheme is based on the monotonic transport 
algorithm and since it employs a staggered mesh, 
the numerical method used in this code is 
spatially second-order accurate. Further details about 
NIRVANA can be found for example in 
De Val-Borro et al. (2006).\\
We adopt computational units in which the mass of 
the central star is $M_\star=1$, the 
gravitational constant is $G=1$ and the cylindrical radius 
$R=1$ corresponds to the initial orbital radius of the planet 
$a_0$. In the following, we report our 
results in units of the initial orbital period of 
the planet $P=2\pi\Omega_0^{-1}$, where 
$\Omega_0=\sqrt{GM_\star/a_0^3}$.\\
For most of the calculations, we employ 
$N_r=128$ radial grid cells uniformly distributed 
between $r_{in}=0.4$  to $r_{out}=2.5$, 
$N_\phi=384$ grid cells in azimuth and $N_\theta=64$ 
meridional grid cells with $\theta$ lying in the range 
$80^\circ$ to $90^\circ$. Some low 
resolution runs using $N_r=64$, $N_\phi=192$ 
and $N_\theta=64$ grid cells have also been 
performed. These are long-term simulations, 
aimed at giving the embedded giant planet 
sufficient time to create a gap in the disk.

\subsection{Initial conditions}

\begin{figure}
\centering
\includegraphics[width=\columnwidth]{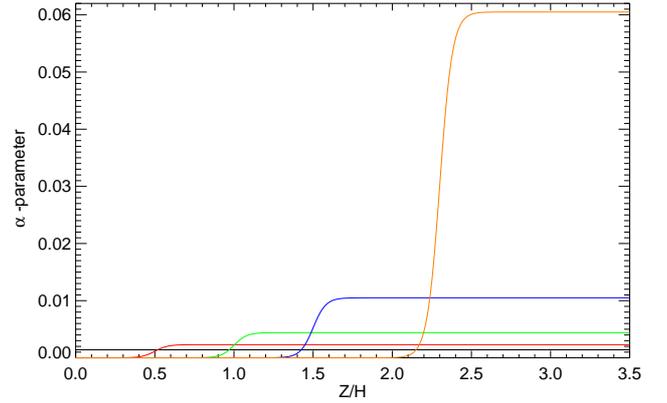}
\caption{This figure shows the $\alpha$-parameter as a
function of the distance from the equatorial plane 
for the different sizes of the dead-zone we
considered, namely for
$\hdz/H=0$ (black line), $\hdz/H=0.5$ (red line), $\hdz/H=1$ (green line), 
$\hdz/H=1.5$ (blue line) and $\hdz/H=2.3$ (orange line). }
\label{fig:alpha}
\end{figure}

In the disk model that we adopt for the simulations presented here, 
the aspect ratio is 
constant and set to $h=H/R=0.05$, which is a typical 
value for protoplanetary disks. The 
initial density profile is chosen such that the 
vertical stratification of the disk follows the 
condition for hydrostatic equilibrium  and is defined by:

\begin{equation}
\rho=\rho_0\left(\frac{R_0}{R}\right)^{3/2}\exp\left(-\frac{Z^2}{2H^2}\right)
\end{equation}

\noindent
where $R$ is the cylindrical disk radius, 
$Z$ is the distance from the midplane and where 
$\rho_0$ and $R_0$ are constants. Consequently, the 
disk surface density $\Sigma=\int \rho dz$ 
can be written as $\Sigma=\Sigma_0 (R_0/R)^{-1/2}$, 
with $\Sigma_0=\sqrt{2\pi}hR_0\rho_0$. 
Here, we set $R_0=1$ and $\rho_0$ is chosen such that the disk contains 
$\sim 3\times 10^{-3}M_\odot$ within the computational 
domain, which corresponds to 
$\sim 0.02\;M_\odot$ interior to 40 AU. \\
The anomalous viscous stress arising from MHD 
turbulence is parameterized using the standard 
'alpha' prescription for the disk viscosity 
$\nu=\alpha c_s H$ (Shakura \& Sunyaev 1973), where $c_s$ 
is the local speed of sound . From 
the results of previous MHD simulations 
(Papaloizou \& Nelson 2003; Fromang \& Nelson 2006), such a prescription is 
expected to provide a reasonable description of the mean 
flow in the disk. It is now understood that
low mass planets are subject to stochastic forcing when 
embedded in turbulent disks (e.g. Nelson \& Papaloizou 2004; Nelson 2005),
and so the adoption of an anomalous viscous stress to
model the effects of turbulence may seem inappropriate
at first sight. But, in this work, we are mainly interested
in the migration and growth of more massive, gap forming
planets where the primary role of turbulence is to supply
gas to the vicinity of the planet through an accretion flow
generated by the turbulent stresses,
and the local interaction between the
planet and turbulent density fluctuations is less important
(e.g. Nelson \& Papaloizou 2003; Papaloizou, Nelson \& Snellgrove 2004).

Although it remains to be proven that for a disk with
a dead-zone, the stresses arising from MHD turbulence 
behave like a locally varying laminar viscosity,
$\alpha$ is chosen here to 
be Z-dependent and such that it can vary from very 
small values in the disk midplane to much higher 
values near the disk surface. More precisely, 
in the calculations we performed, the 
vertical profile adopted for $\alpha$ was as follows:

\begin{equation}
\alpha(Z)=\frac{1}{2}\left\{(\alpha_{AZ}-\alpha_{DZ})\tanh\left(\frac{Z-\hdz}{\Delta}\right)+
\alpha_{AZ}+
\alpha_{DZ} \right\},
\label{oldalpha}
\end{equation}

\noindent
where  $\hdz$ is the size of the dead-zone and where 
$\alpha_{AZ}$ (resp. $\alpha_{DZ}$) is the 
value for the $\alpha$-parameter in the active (resp. dead) 
region. In the previous equation, 
$\Delta$ denotes the size of the transition from 
the dead to the active zone and is set to 
$\Delta=0.08H$ in the work presented here. 
In order to study the influence of the dead-zone 
size on the evolution of giant planets, we 
have considered different values for $H_{DZ}$ 
lying in the range $0$ to $2.3 \;H$. In the case 
where $\hdz=0$, $\alpha$ is constant in the 
entire disk and is set to $\alpha_0=1.4\times 10^{-3}$, 
which corresponds to a constant 
accretion rate through the disk 
$\dot M_0=3\pi\alpha_0\Sigma_0 h^2$. For other values of 
$H_{DZ}$, we adopt a similar value for the vertically 
averaged mass flow through the layered 
disk $\dot M=3\pi\int \nu(z)\;\rho dz=\dot M_0$ and we 
calculate the value for $\alpha_{AZ}$ 
accordingly, assuming that the residual viscosity in the dead-zone is 
$\alpha_{DZ}=1.0\times 10^{-7}$. For the different 
values of $H_{DZ}$ we considered, the 
corresponding value for $\alpha_{AZ}$ can be found in Table 1 
 and Fig. \ref{fig:alpha} displays the corresponding $\alpha$-parameter
as a function of the distance from the disk equatorial plane .\\

The planet is initially placed on a circular orbit at 
$a=1$, and our main interest is to examine the orbital evolution of
planets with masses $m_p=3\times 10^{-4}$ and $10^{-3}\;M_*$,
corresponding to Saturn-mass ($M_S$) and
 Jupiter-mass ($M_J$) planets, respectively. 
In order to conserve mass as the planet grows and opens a gap,
we proceed in two steps:\\
i) we consider a  $30\; M_\oplus$ protoplanet held at 
$a=1$ on a fixed circular orbit. This 
body is allowed to accrete gas from the disk on a 
timescale corresponding to $\sim 1500$ 
orbits. Due to the very long run times required for the 
planet mass to reach $1\;M_J$, we 
have performed, for this step, low-resolution 
simulations using $N_r=64$, $N_\phi=192$ and 
$N_\theta=64$ grid cells. \\
ii) once the planet mass has reached $1\;M_S$ or $1\;M_J$, 
we restart the calculations, but 
with the planet being allowed to migrate due to its 
interaction with the disk, and accretion being switched off. For this second step, 
we have performed higher resolution simulations using 
$N_r=128$, $N_\phi=384$ and $N_\theta=64$ 
grid cells. The new values for the disk physical 
quantities were computed from the old ones 
using bi-linear interpolation. 
  
\begin{table}
\caption{Dead-zone parameters for the different models considered.}
\label{table1}
\centering{}
\begin{tabular} {c c c}
\hline \hline
$H_{DZ}/H$ & $\alpha_{AZ}$ & $\alpha_{DZ}$ \\
\hline
$0.5$ & $2.3\times 10^{-3}$ & $1\times 10^{-7}$ \\
$1$   & $4.4 \times 10^{-3}$ & $1\times 10^{-7}$ \\
$1.5$ & $1.05 \times 10^{-2}$ & $1\times 10^{-7}$ \\
$2.3$ & $6.5 \times 10^{-2}$ & $1\times 10^{-7}$\\
\hline
\end{tabular}
\end{table}

\subsection{Boundary conditions}

At the inner edge of the computational domain, we model 
accretion onto the central star by 
using a `viscous' boundary condition, for which we set the 
radial velocity in the 
innermost cells to $v_r=\beta v_{visc}(r_{in})$, where 
$v_{visc}(r_{in}) =-3\nu/2r_{in}$ is the 
typical disk inward drift velocity due to 
viscous diffusion, and $\beta$ is a free 
parameter which is set to $\beta=5$ in this work (Pierens \& Nelson 2008). 
At the outer edge of 
the computational domain, and at the lower meridional boundary as well, 
we prevent mass loss from 
the disk by employing reflecting boundary conditions.
Assuming that the disk is symmetric with 
respect to its midplane, we also impose a symmetry boundary condition at the upper 
meridional boundary, which corresponds to the disk midplane. 
Consequently, we  consider 
only the upper half of the disk in the simulations described below.

\section{Results}

\subsection{Effect of the presence of a dead-zone on planetary growth}
\label{part1}

\begin{figure}
\centering
\includegraphics[width=0.95\columnwidth]{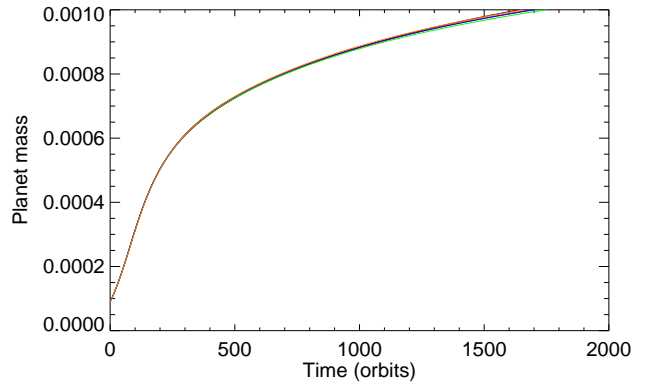}
\caption{ This figure shows the evolution of the 
planet mass as a function of time for the 
different sizes of the dead-zone we consider, namely for
$\hdz/H=0$ (black line), $\hdz/H=0.5$ (red line), $\hdz/H=1$ (green line), 
$\hdz/H=1.5$ (blue line) and $\hdz/H=2.3$ (orange line).}
\label{massvstime}
\end{figure}

For each value of $\hdz$ we consider, we have performed a simulation 
in which a $30\;\mearth$ 
planet held on a circular orbit can accrete gas from the disk until its mass reaches 
$1\;M_J$ .  In the simulations presented here accretion 
is modelled by removing at each 
time-step a fraction of the gas located inside the Roche 
lobe of the planet and then adding the 
corresponding amount of matter to the mass of the planet 
(e.g. Kley 1999; Nelson et al. 2000). 
The e-folding time for gas accretion was chosen to 
be the orbital period of the planet, which 
corresponds to the maximum rate at which the planet can 
accrete gas (Kley 1999). The choice of a $30 \; \mearth$
protoplanet to act as a seed onto which gas can accrete
is broadly consistent with evolutionary models of gas giant 
planets forming in protoplanetary 
disks, which suggest that rapid gas accretion occurs 
once the planet mass reaches $30-40\;\mearth$ (Papaloizou \& Nelson 2005).\\
The time evolution of the planet mass for each model is 
displayed in Fig. \ref{massvstime}. 
We see that the evolution of the planet mass depends only weakly on the size of the 
dead-zone and proceeds similarly in each case. The early evolution involves rapid gas accretion 
with the planet growing to become a Saturn-mass planet in 
$\sim 100$ orbits. As the planet mass 
increases, non-linear effects can tidally truncate the disk, leading to a 
decrease in the accretion rate such that it takes 
$\sim  1500$ orbits for a Saturn-mass 
planet to reach a Jovian mass. \\
 
\noindent  
It is worth noting, however, that although at earlier times 
the accretion rate should not depend 
strongly on the size of the dead-zone,  since
the planet is able to accrete material which is
on neighbouring orbits, we expect this to 
be not true from the time the gravitational
torque is largely deposited in the disk in the near-vicinity of the
planet. This occurs when the thermal criterion for gap opening is satisfied
(Lin \& Papaloizou 1986; Crida et al. 2006), namely when the radius of the Hill 
sphere exceeds the disk semi-thickness:  

\begin{equation}
a\left(\frac{m_p}{3M_*}\right)^{1/3} > H.
\end{equation}

\noindent In the simulations presented here, this happens roughly when 
$m_p \ge 3\times 10^{-4}$. From this time, the planet may be 
able to open a gap in the disk 
provided that the gravitational torques overwhelm the viscous 
forces (Bryden et al. 1999; 
Papaloizou et al. 2006), a condition which can be expressed as:

\begin{equation}
\frac{m_p}{M_*} > \frac{40\nu}{a^2\Omega_p},
\label{viscous}
\end{equation}

\begin{figure*}
\centering
\includegraphics[width=0.9\columnwidth]{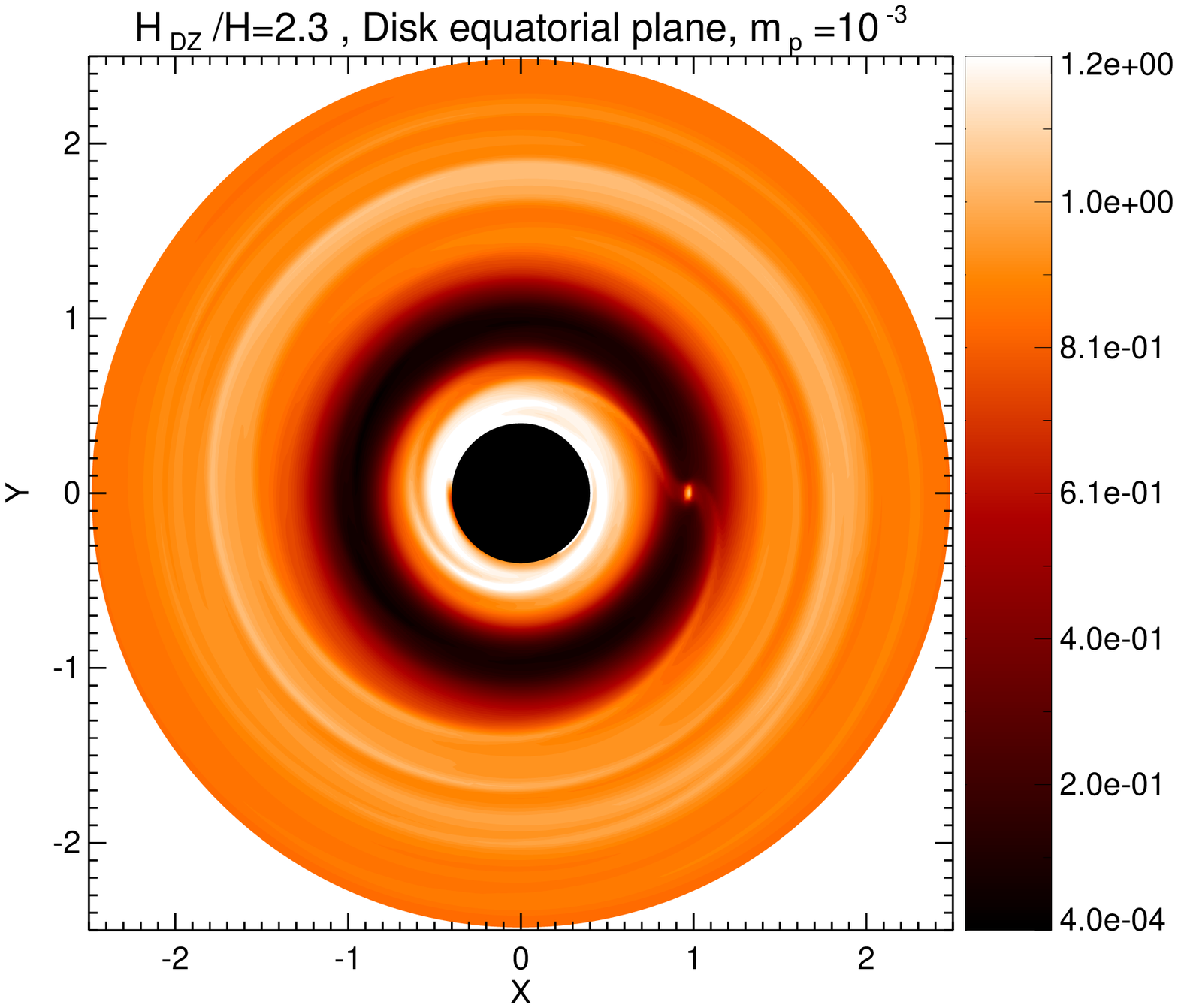}
\includegraphics[width=0.9\columnwidth]{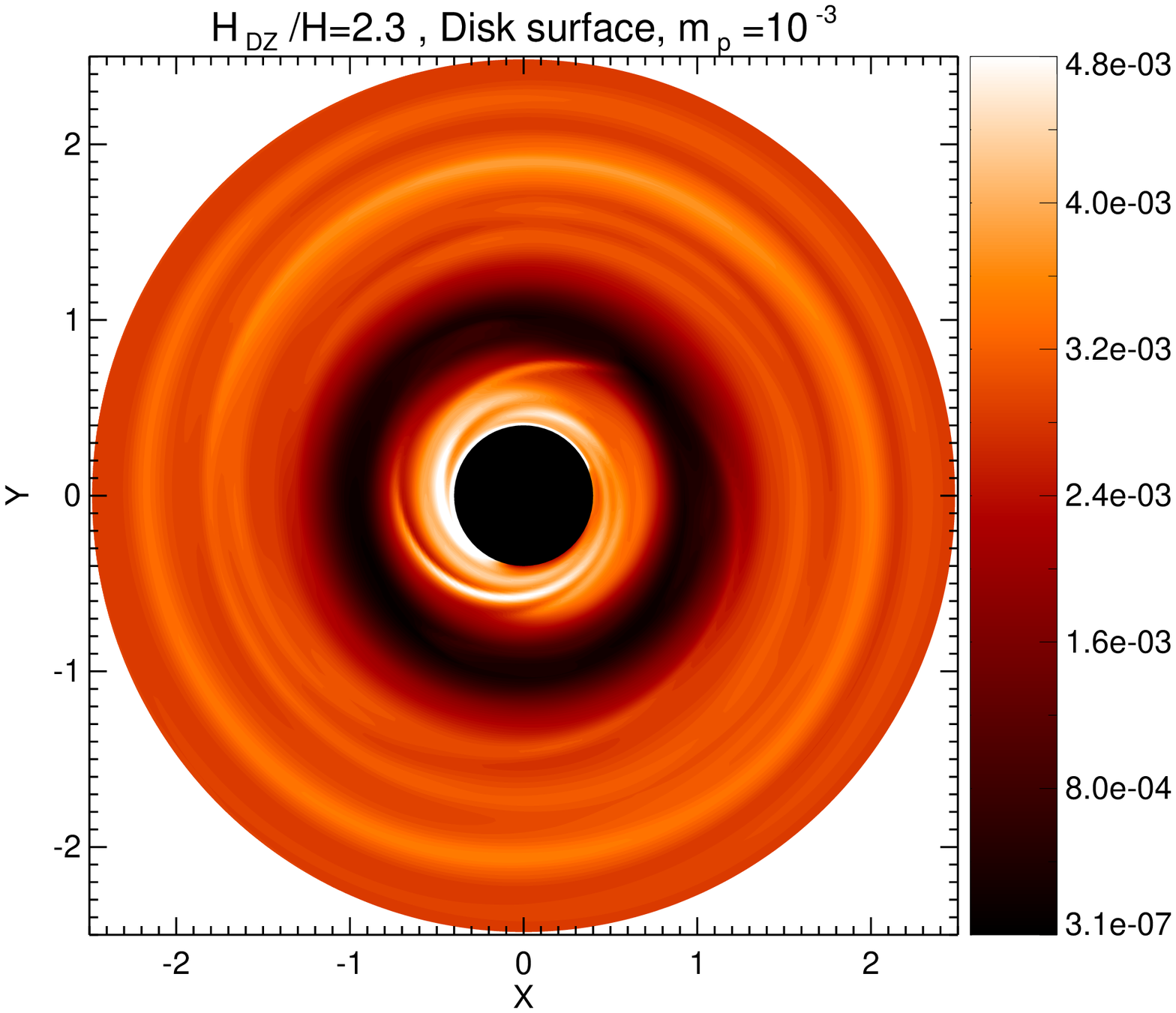}
\caption{{\it Left panel}: this figure shows a snapshot of 
the disk density in 
the equatorial plane for $\hdz/H=2.3$. Here, 
a planet held on a circular orbit at $a=1$ and with initial mass 
$m_p=30\;M_\oplus$ has
grown to become a Jupiter-mass planet. 
{\it Right panel}: same but at the disk upper surface.   }
\label{jupiter2}
\end{figure*}
\begin{figure}
\includegraphics[width=\columnwidth]{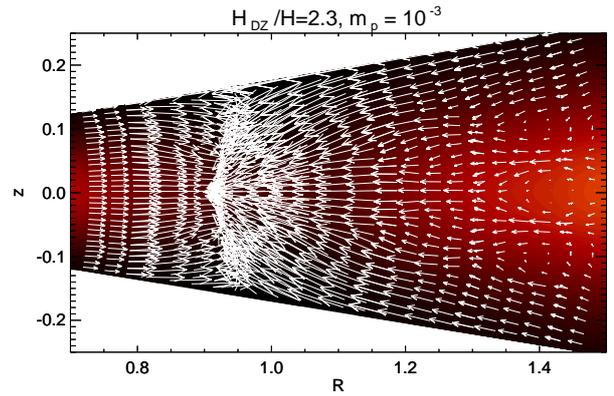}
\caption{This figure shows, for $\hdz/H=2.3$,  details of the flow in the neighbourhood of the 
accreting planet when its mass  has reached $1\;M_J$. The velocity field is represented
in the $(R,z)$ plane and for ${0.6\le R \le 1.5}$.}
\label{velocity}
\end{figure}


\noindent
where $\Omega_p$ is the angular velocity of the planet.
Since the value for the viscosity in the active region is 
correlated with the size of the 
dead-zone, the previous equation shows that the gap structure 
and consequently the accretion 
rate onto the planet may depend strongly on 
$\hdz$ for $m_p\ge 3\times 10^{-4}$. Indeed, 
considering for instance the case $m_p=1\;M_J$, 
Eq. \ref{viscous} predicts that gap opening 
should occur for $\alpha \le 0.01$. Therefore, in the 
cases where $\hdz/H=1.5$ and 
$\hdz/H=2.3$, we expect a $1\;M_J$ planet to be able 
to open a gap only in the dead-zone, possibly
leaving the gas flowing over the top of the planet in the active region
if the accretion flow there is rapid enough. 
Snapshots of the disk density in the equatorial plane,
and at the disk surface, are depicted in Fig. \ref{jupiter2}. 
These plots correspond to a time when the mass of the accreting 
protoplanet has reached $1\;M_J$ and are for a 
model in which $\hdz=2.3H$. These reveal that the planet 
truncates the disk throughout its 
vertical extent,  despite the high value for the viscosity 
in the active region. As can be 
seen in Fig. \ref{velocity}, such a process arises because 
the planet is able to pull some of 
the gas from the live-zone down in toward the midplane as 
the latter tries to flow over the top 
of the planet. This suggests that both the gap structure and the 
accretion rate onto the 
planet do not depend strongly on the presence of a dead-zone, but rather on the 
vertically-integrated accretion rate of gas through the disk. In that case, we would 
expect the results from these 3D runs to not significantly differ 
from 2D calculations for a 
given value of the accretion rate $\dot M$. \\

\begin{figure}
\centering
\includegraphics[width=0.95\columnwidth]{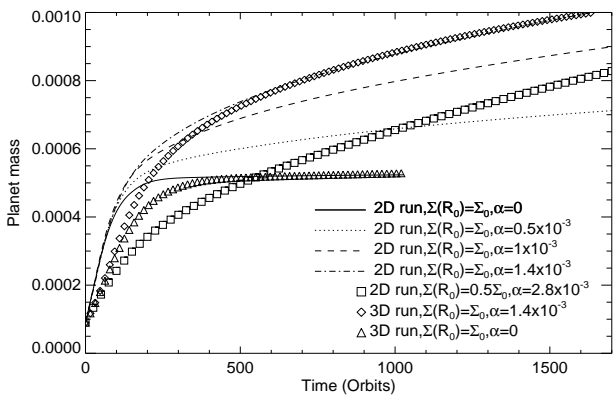}
\caption{This figure shows the evolution of the planet mass as a 
function of time for 2D simulations 
in which the $\alpha$-parameter was varied from 
$\alpha=0$ to $\alpha=2.8\times 10^{-3}$. Diamonds 
display for comparison the results of a 3D run with 
$\hdz=0$ and $\alpha=1.4\times 10^{-3}$ whereas 
triangles display the results of a 3D run with $\alpha=0$.} 
\label{2d3daccret}
\end{figure}

To investigate this issue in more detail, we have 
performed a suite of 2D runs with varying values of $\alpha$ from $\alpha=0$ to 
$\alpha=1.4\times 10^{-3}$ and for which the initial surface density
at the planet position is $\Sigma(R_0)=\Sigma_0$. An additional 2D 
calculation with
$\alpha=2.8\times 10^{-3}$  and $\Sigma(R_0)=0.5\Sigma_0$ has also
been performed. The results of these 
calculations are shown in Fig. \ref{2d3daccret}. 
Also displayed are the results of two 3D simulations corresponding to: 
i) a fully active disk with $\alpha=1.4 \times 10^{-3}$ 
throughout; ii) a dead disk with 
$\alpha=0$ everywhere. As expected, 2D runs 
with $\alpha \ne 0$ predict that the protoplanet 
grows faster as the viscosity increases.  The run with 
$\alpha=2.8\times 10^{-3}$ shows slower growth at earlier times 
since the surface density
is smaller in that case. However, once the planet opens a gap in the disk,
 it is clear that the growth rate becomes similar to that obtained 
from the run with ${\alpha=1.4\times10^{-3}}$ and for which the value
 for the mass accretion
rate through the disk is the same. In the case where 
$\alpha=0$, viscous supply of gas 
through the disk is inhibited and the planet mass can saturate once the feeding zone 
is empty of gas, which occurs when $m_p\sim 0.5 M_J$ 
for the disk parameters used in this work. 
Clearly, for both $\alpha=0$ and $\alpha=1.4\times 10^{-3}$, 
there is a good agreement between 
2D and 3D calculations. This confirms that the accretion rate 
onto a  gap-opening planet does not 
depend strongly on the details of the flow in the $(R,z)$ plane, but only on the 
vertically-averaged accretion rate through the disk. 
  
\subsection{Effect of a dead-zone on the orbital evolution of Saturn-mass planets }

\begin{figure}
\centering
\includegraphics[width=\columnwidth]{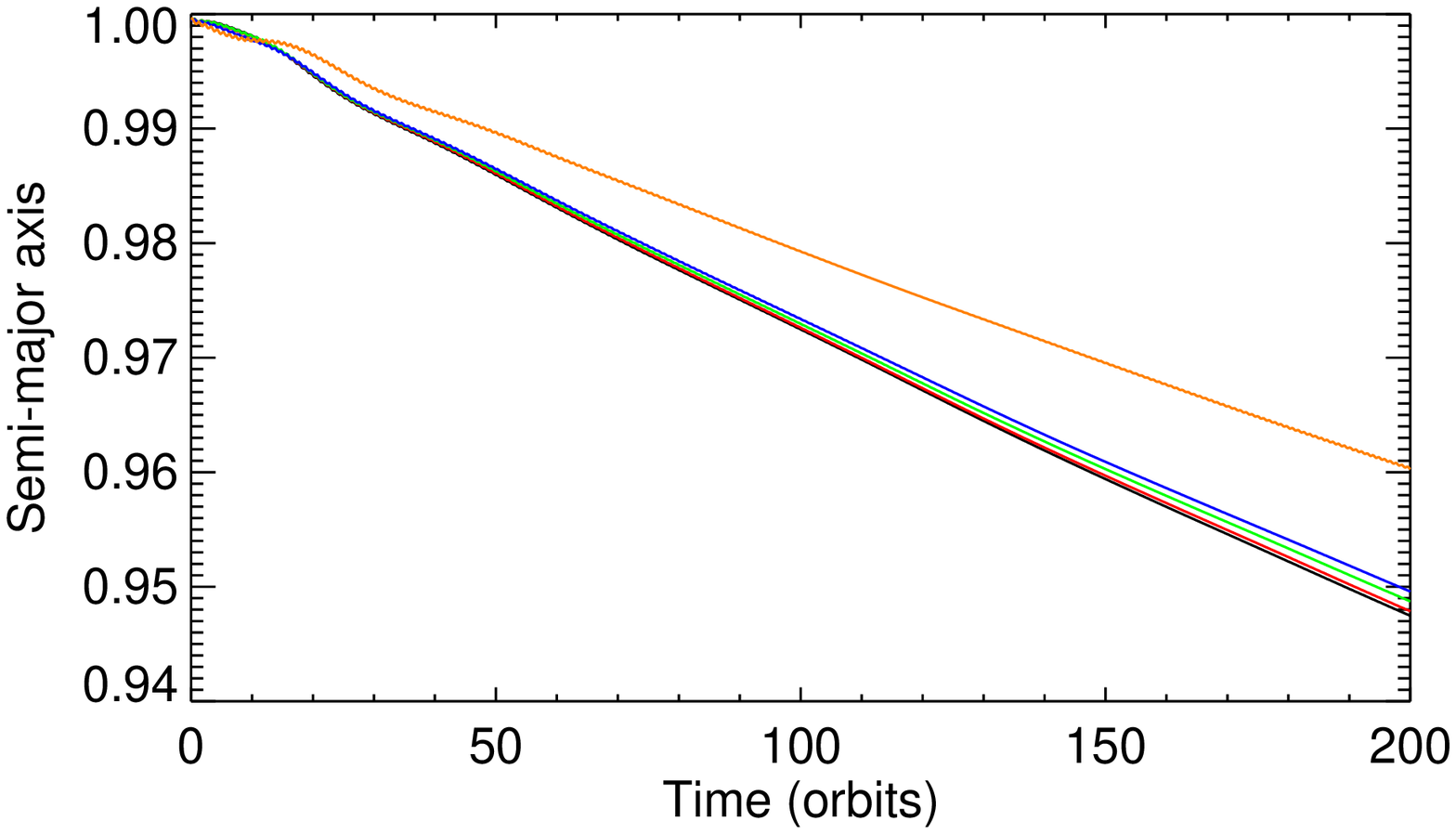}
\includegraphics[width=\columnwidth]{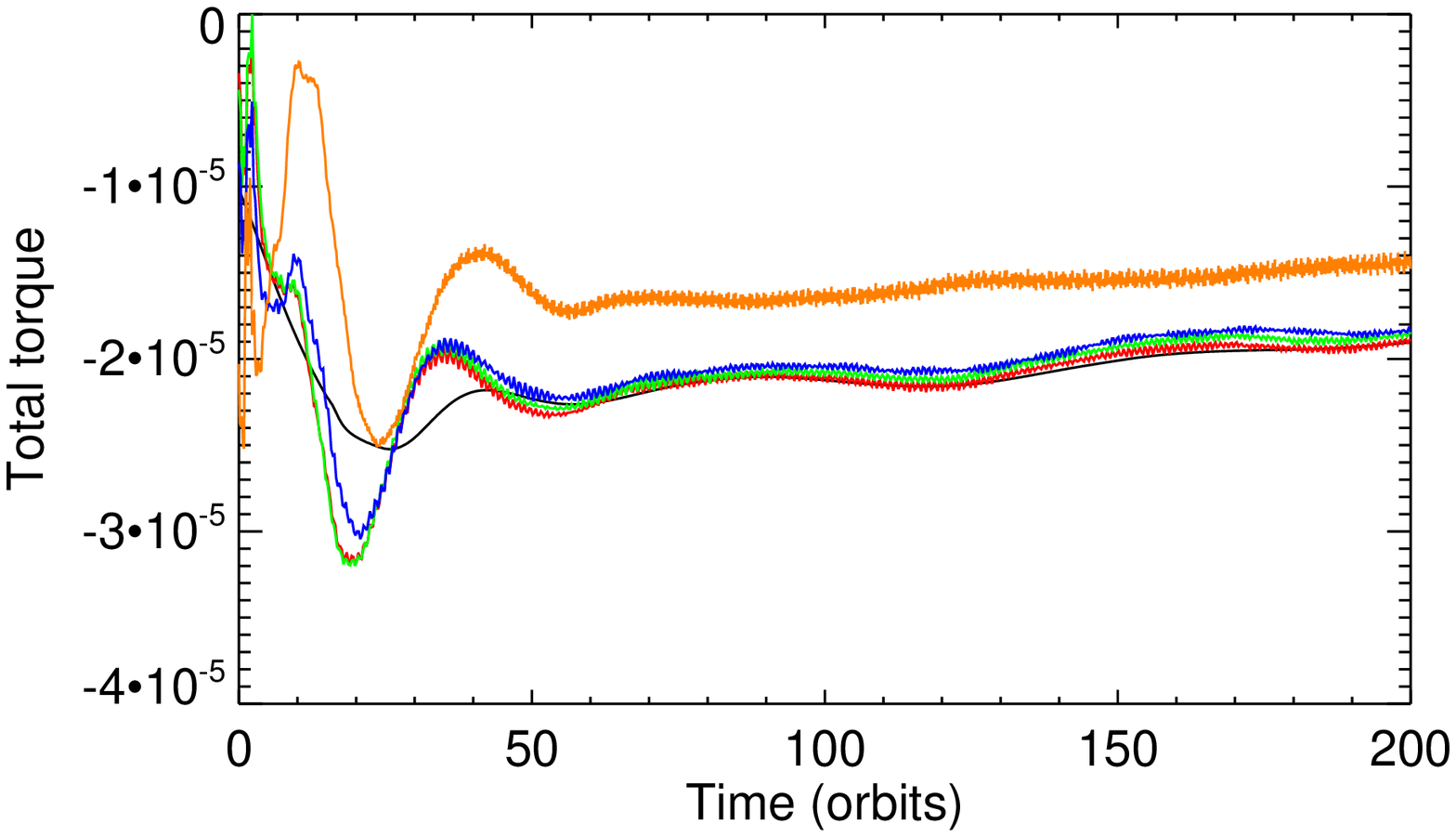}
\caption{{\it Upper panel:} this figure shows the evolution of Saturn's 
semi-major axis for the 
different sizes of the dead-zone we consider, 
namely for $\hdz/H=0$ (black line), $\hdz/H=0.5$ 
(red line), $\hdz/H=1$ (green line), 
$\hdz/H=1.5$ (blue line) and $\hdz/H=2.3$ (orange line). 
{\it Lower panel:} this figure shows the corresponding total torque exerted on the planet.}
\label{saturn}
\end{figure}

\begin{figure*}
\centering
\includegraphics[width=0.9\columnwidth]{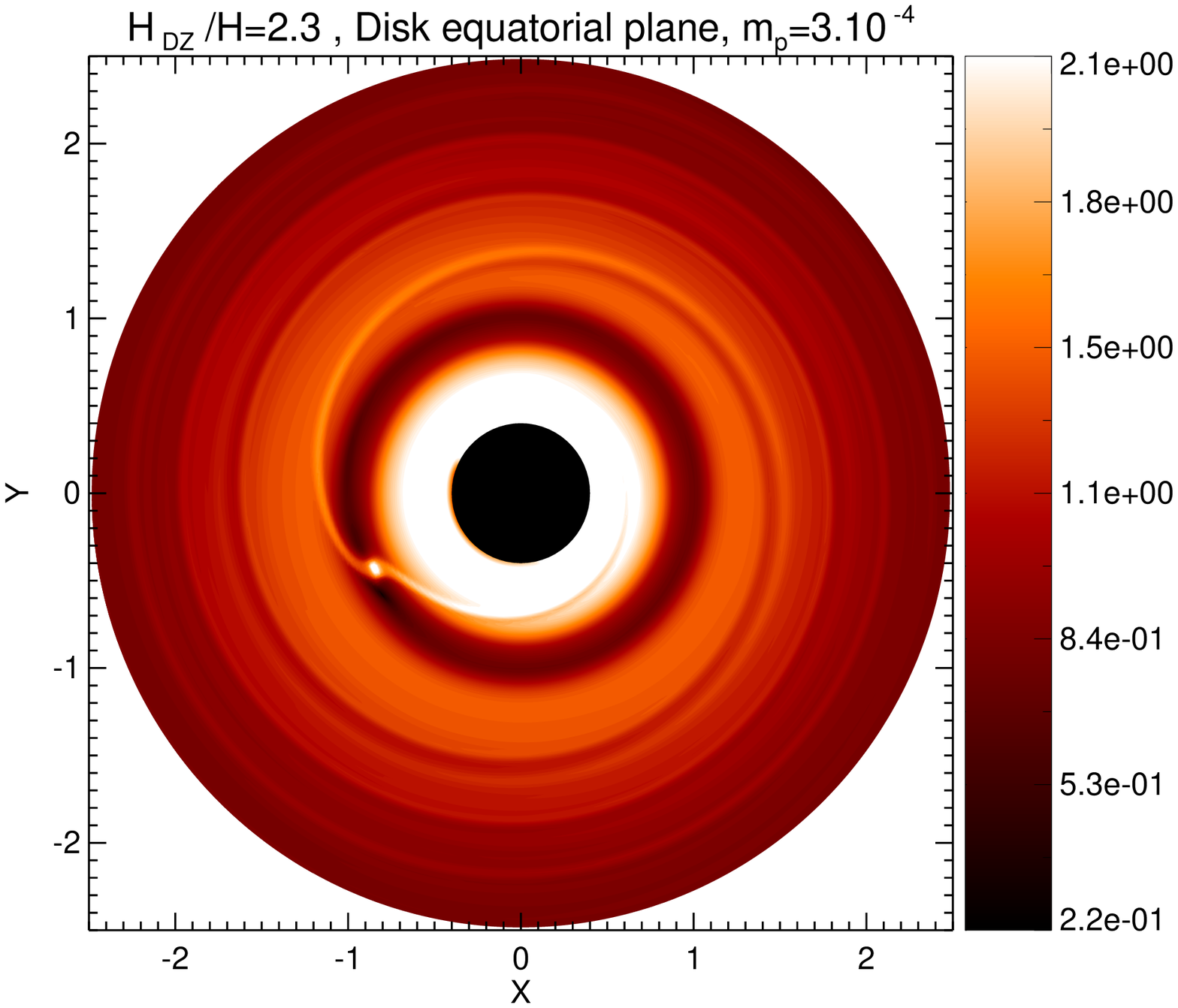}
\includegraphics[width=0.9\columnwidth]{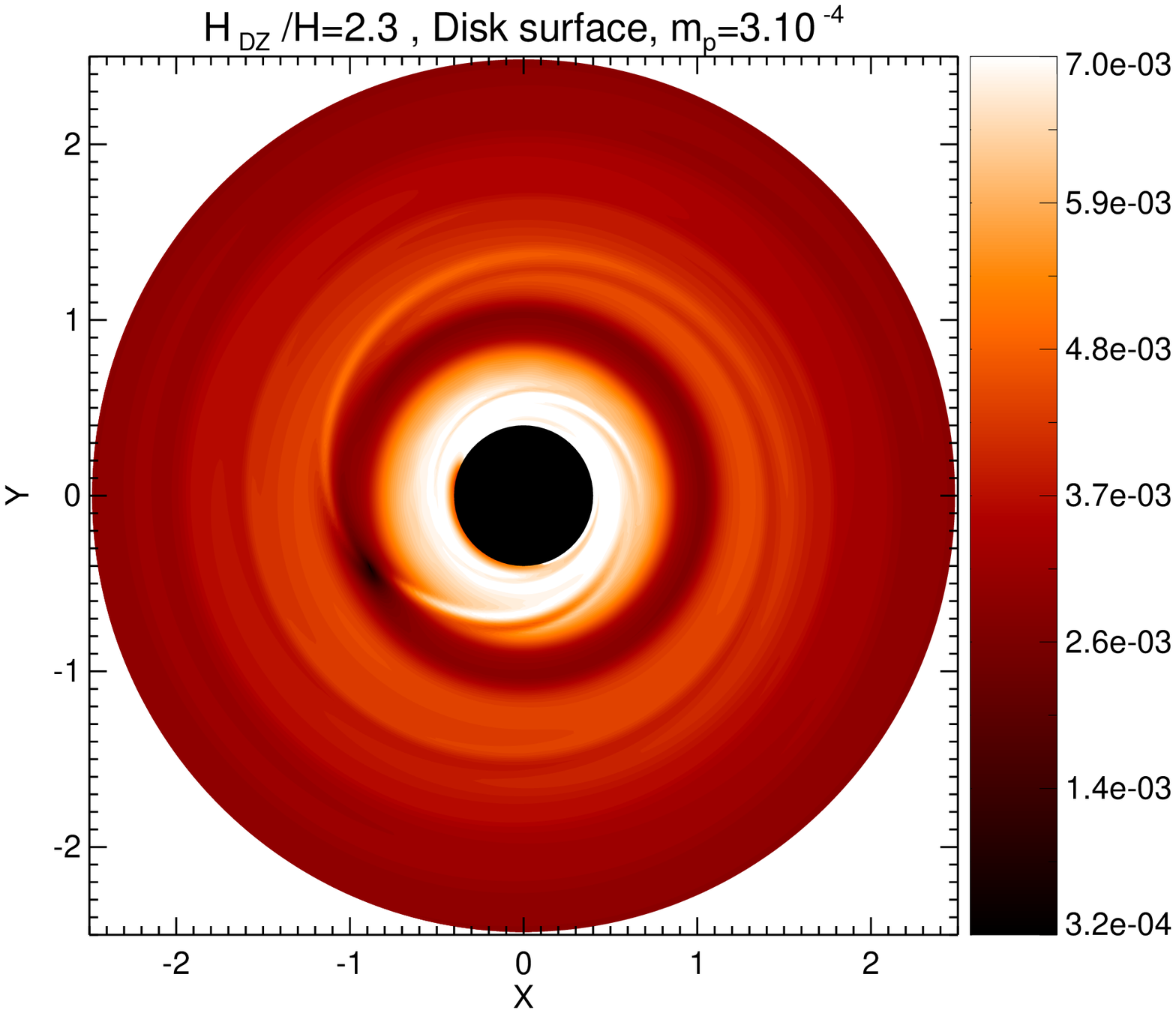}
\caption{{\it Left panet}: this figure shows,for $\hdz/H=2.3$, a 
snapshot of the disk density in 
the equatorial plane at a time where a planet held 
on a circular orbit at $R=1$ and with initial 
mass $m_p=30\;M_\oplus$ has grown to become a 
Saturn-mass planet. {\it Right panel}: same but at 
the disk surface.}
\label{saturn2}
\end{figure*}

In order to study the evolution of a Saturn-mass planet 
embedded in a layered protoplanetary 
disk, we restarted the simulations described in the 
previous section when the mass 
of the accreting planet had reached the relevant value. 
We then released the planet and let it 
evolve freely under the action of disk torques.\\
The evolution of Saturn's semi-major axis for the different 
models is shown in the upper panel 
of Fig. \ref{saturn}. For each value of $\hdz/H$, Saturn does not undergo 
runaway migration 
(Masset \& Papaloizou 2003) because the disk is not massive enough.
Instead, the migration rate is
intermediate between the Type II and Type I regimes. 
This arises because the mass of Saturn is insufficient
to enable a clean gap to form in the disk, as illustrated in Fig. \ref{saturn2}.
This figure displays, for the model with $\hdz/H=2.3$, a snapshot 
of the Saturn-induced gap structure in both the
equatorial plane and at the disk surface. It is worthwhile noticing that 
gap opening occurs in the live-zone, despite the high value for 
the viscosity there, indicating 
that both Jupiter and Saturn can pull gas from the active region down 
toward the disk midplane.\\
For models with $\hdz/H \le 1.5$, the evolution of the system 
proceeds quite similarly, with a 
slight tendency  for the migration rate to decrease 
as the size of the dead-zone increases. For 
the calculation with $\hdz/H=2.3$ however, we see that in 
$\sim 200$ orbits the semi-major axis 
has decreased by an amount that is nearly $20\%$ less than for the other cases. 
This can be confirmed by examining 
the total disk torques exerted on the planet, which are displayed for each 
model in the lower panel of Fig. \ref{saturn}.
It is well known that the total tidal torque, $\Gamma$, 
exerted on the planet can be written as the sum of a 
differential Lindblad torque, $\Gamma_L$, 
and a total corotation torque, $\Gamma_C$ (Tanaka \& Ward 2002). 
In contrast to $\Gamma_L$, which is 
 basically independent of viscosity 
(Meyer-Vernet \& Sicardy 1987; Papaloizou \& Lin 1984), 
$\Gamma_C$  depends strongly on $\nu$ and can saturate 
in the low-viscosity limit. In the 
present case, $\Gamma_C$ can be decomposed as:

\begin{equation}
\Gamma_C=\Gamma_C^{DZ}+\Gamma_C^{AZ},
\end{equation}

\noindent
where $\Gamma_C^{DZ}$ (resp. $\Gamma_C^{AZ}$) is the corotation torque 
due to the librating 
fluid elements which originate from the dead (resp. active) region. 
In the dead-zone, 
$\Gamma_C^{DZ}$ is expected to saturate on a timescale corresponding to 
the outermost horseshoe turnover time (Masset 2002)

\begin{equation}
\tau_{HS}=\frac{8\pi a}{3x_s}\Omega_p^{-1}.
\end{equation}

\begin{figure}
\centering
\includegraphics[width=0.95\columnwidth]{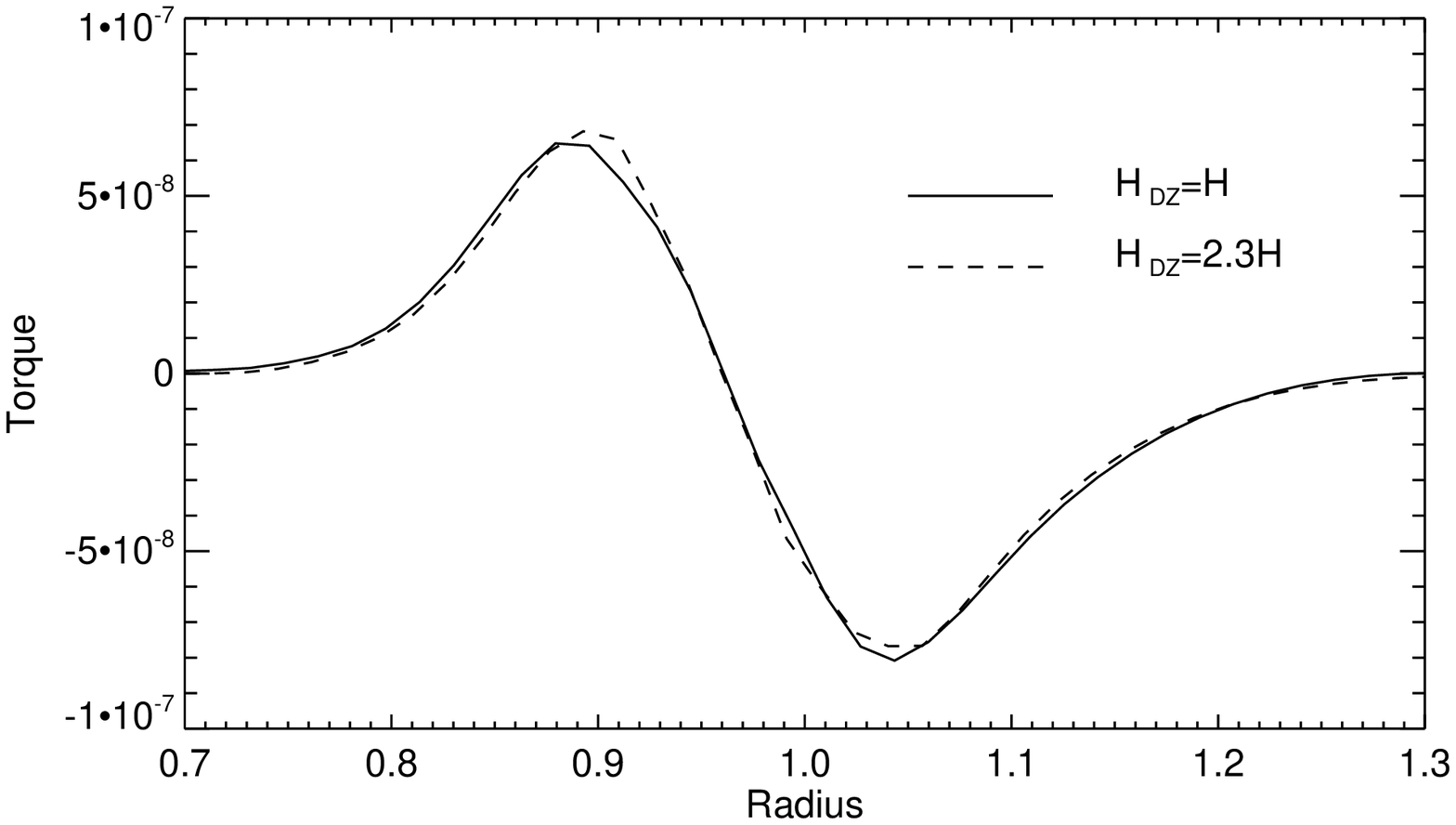}
\includegraphics[width=0.95\columnwidth]{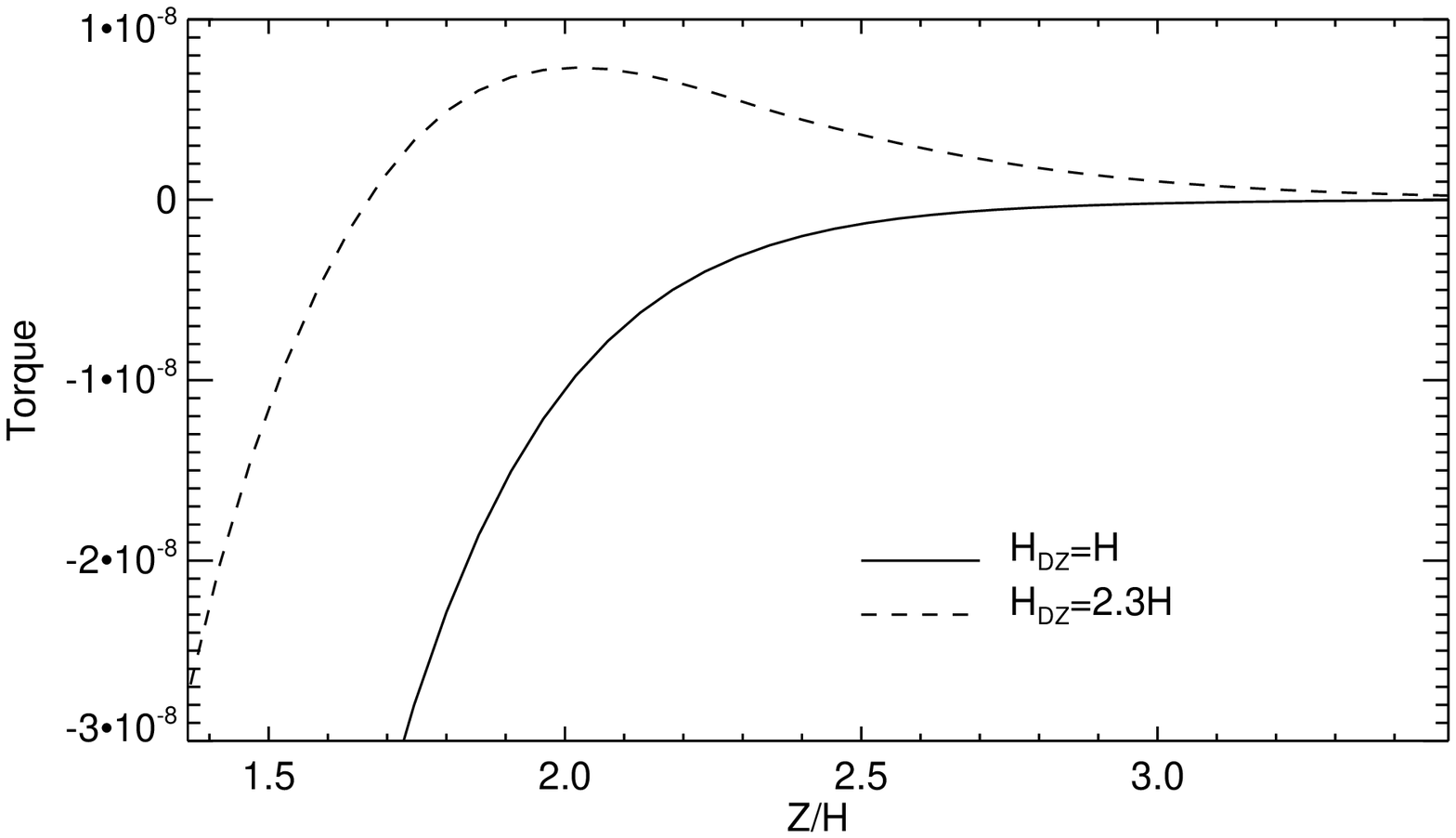}
\caption{{\it Upper panel:} this figure shows the radial distribution 
of the torque for the runs with $\hdz=H$ and $\hdz/H=2.3$. The planet is 
here located at $a=0.96$.   
{\it Lower panel:} this figure shows the vertical distribution of the torque
in the upper disk layers.}
\label{torquedistrib}
\end{figure}

\noindent 
In the previous equation, $x_s$ is the horseshoe zone width which 
can be approximated by 
$x_s=2.45a(q/3)^{1/3}$ (Masset et al. 2006), 
where $q$ is the planet to central star 
mass ratio. Here, this gives $\tau_{HS}\sim 71\; \Omega_p^{-1}$, 
which is consistent with the 
timescale over which the total torques reach a constant 
value in Fig. \ref{saturn}. In the 
active region, however, we would expect the corotation torque 
 $\Gamma_C^{AZ}$ there to saturate only if the 
viscous time-scale $\tau_v=x_s^2/3\nu$ across the 
horseshoe region exceeds $\tau_{HS}$. 
Assuming that $x_s$ is constant over the vertical extent of the disk,
and for the disk parameters 
used in this work, this suggests that the corotation torque saturates 
in the active region 
provided that $\alpha_a\le 2.5\times 10^{-2}$. Therefore, 
we would expect the corotation torque 
to saturate in models with $\hdz/H\le 1.5$. For the 
calculation with $\hdz/H=2.3$ however, we 
would expect the corotation torque originating from 
the live-zone to be positive since for our 
disk model, the disk surface density decreases 
$R^{-1/2}$. Such a scenario is consistent with 
the fact that the total torques exerted on Saturn do 
not depend on the size of the dead-zone 
for $\hdz/H=1.5$, and are weaker in the case where 
$\hdz/H=2.3$ (see Fig. \ref{saturn}).  For the 
run with $\hdz=2.3H$, the computed mass of gas 
material located inside the corotation region 
is $\sim 8\times 10^{-5}$ in our units with about 
$3\; \%$ of this mass located in the active 
zone, which appears to be sufficient to 
provide a non negligible contribution to the 
total corotation torque exerted on the planet. In order to
clearly demonstrate that the difference in the torques arise 
from the  viscously-evolving corotation region, we show in the upper panel 
of Fig. \ref{torquedistrib}  
the radial torque distribution $\Gamma(r)$  for the runs 
with $\hdz=H$ and $\hdz=2.3H$  and for a planet located at
$a=0.96$.
 In comparison with the simulation in which $\hdz=H$, the run 
with $\hdz=2.3H$ exhibits a higher positive bump at $r\sim 0.89$
and a smaller negative bump at $r\sim 1.04$, which therefore 
corresponds to a slightly smaller
differential Lindblad torque. For the run with $\hdz=2.3H$, the 
positive torque excess for $0.9<r<0.94$ can be explained by 
looking at the vertical torque distribution 
$\Gamma(z)$ exerted by the disk region located between $a-x_s$ 
and $a+x_s$ and which is displayed in the lower panel of 
Fig. \ref{torquedistrib}. For this simulation, the torque 
becomes positive beyond $Z > 1.7H$, which is clearly not the case
for  the run with $\hdz=H$. This unambiguously confirms that
the corotation torque exerted by the viscously-evolving
layers is unsaturated in the
run with $\hdz=2.3H$ and tends to slow down migration.

\subsection{Orbital evolution of Jupiter-mass planets}

\begin{figure}
\centering
\includegraphics[width=\columnwidth]{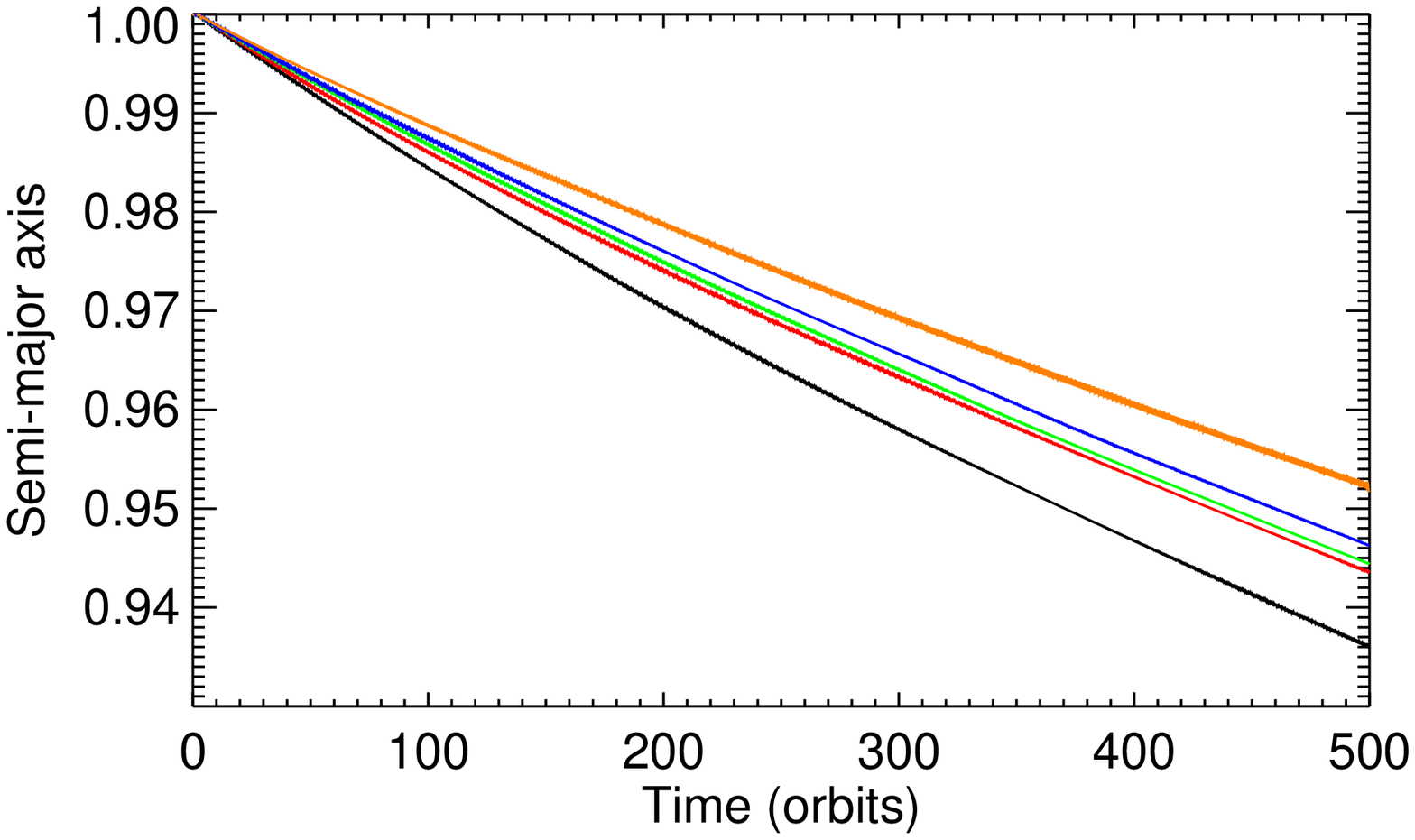}
\includegraphics[width=\columnwidth]{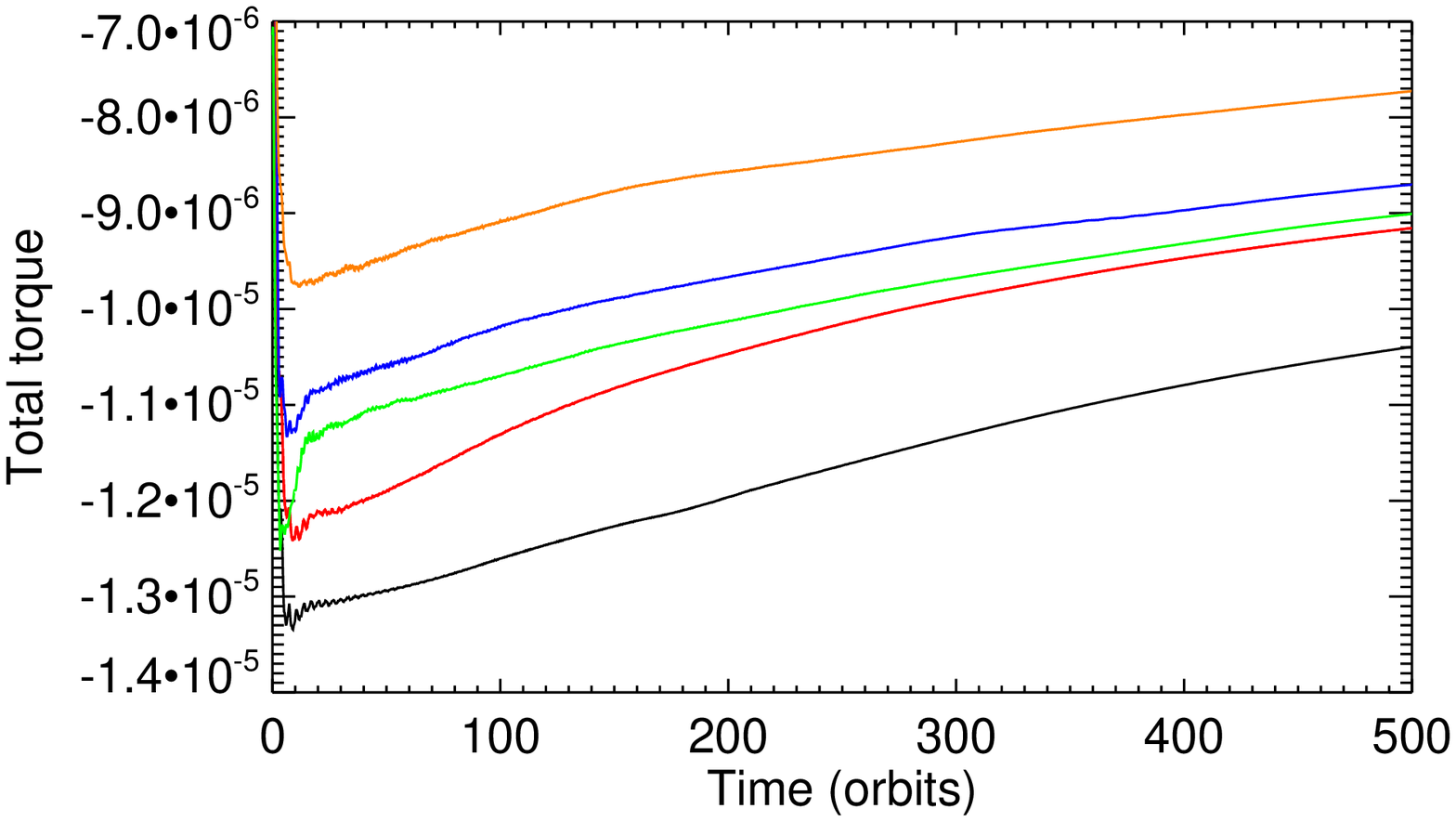}
\caption{{\it Upper panel}: this figure shows the evolution of 
Jupiter's semi-major axis for the 
different sizes of the dead-zone we consider, 
namely for $\hdz/H=0$ (black line), $\hdz/H=0.5$ 
(red line), $\hdz/H=1$ (green line), 
$\hdz/H=1.5$ (blue line) and $\hdz/H=2.3$ (orange line). 
{\it Lower panel:} this figure shows the corresponding total torque exerted on Jupiter.}
\label{jupiter}
\end{figure}

\begin{figure}
\centering
\includegraphics[width=\columnwidth]{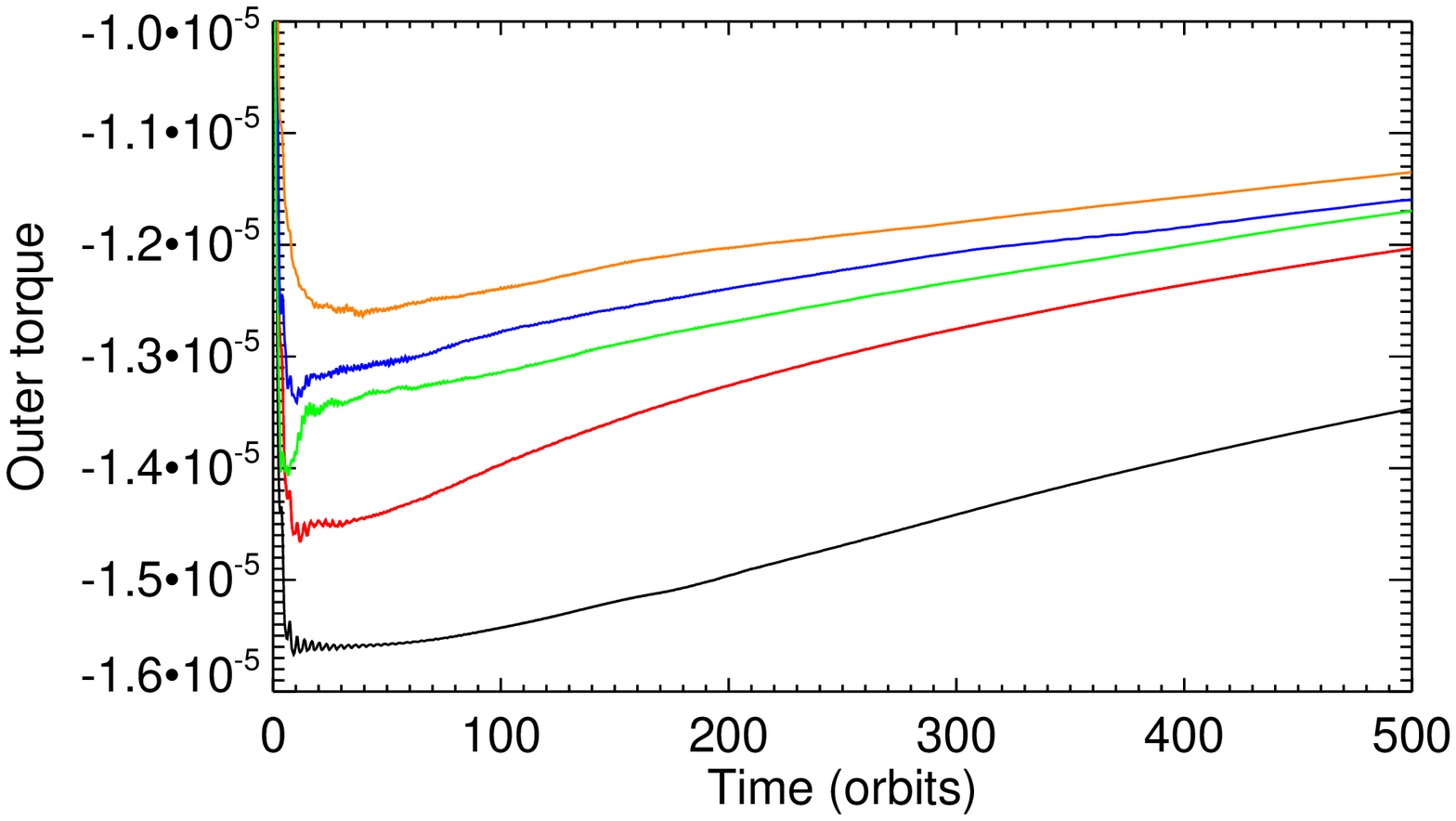}
\includegraphics[width=\columnwidth]{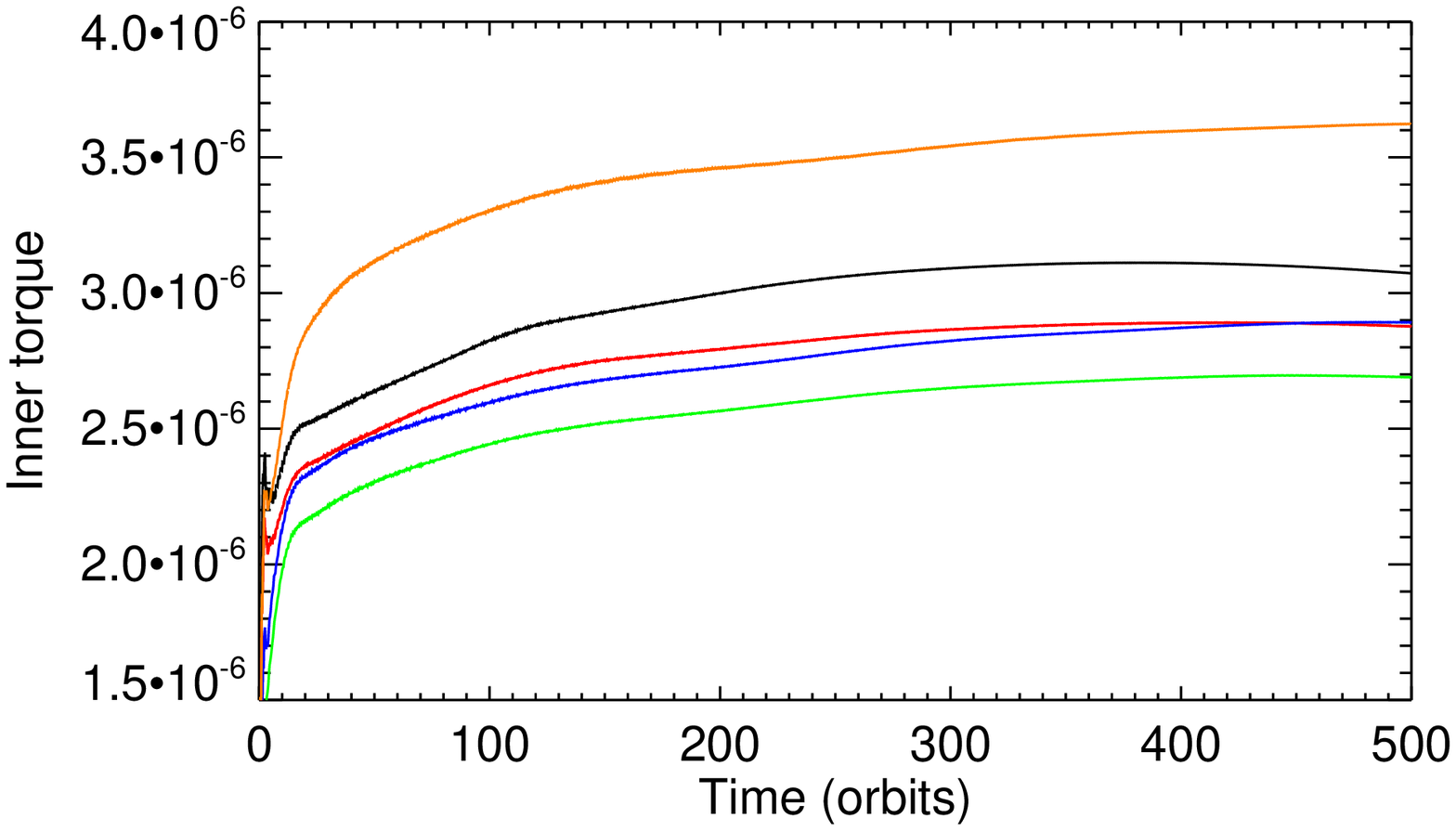}
\caption{{\it Upper panel:} this figure shows the time evolution of the outer torque exerted on 
Jupiter, for the different sizes of the dead-zone we consider,namely for $\hdz/H=0$ (black line), 
$\hdz/H=0.5$ (red line), $\hdz/H=1$ (green line), $\hdz/H=1.5$ (blue line) and $\hdz/H=2.3$ 
(orange line). {\it Lower panel:} same but for the inner torque. }  
\label{torques-jupiter}
\end{figure}

\begin{figure}
\centering
\includegraphics[width=\columnwidth]{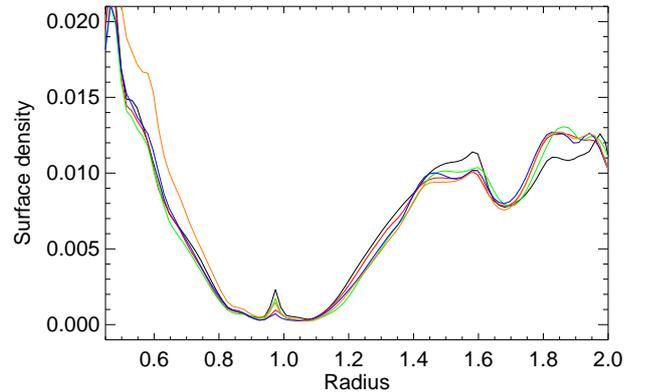}
\caption{This figure shows, for the simulations in which $m_p=10^{-3}$,
 the disk surface density as a function of radius and for an azimuthal 
position corresponding to that of the planet. Different sizes of the dead-zone are considered, 
namely $\hdz/H=0$ (black line), $\hdz/H=0.5$ (red line), $\hdz/H=1$ (green line), $\hdz/H=1.5$ 
(blue line) and $\hdz/H=2.3$ (orange line). }
\label{surface-density}  
\end{figure}

We now turn to the question of how the orbit of a Jupiter-mass 
planet evolves in a dead-zone. Here again, to 
address this issue, we restarted the simulations presented 
in Sect. \ref{part1} once the mass 
of the accreting protoplanet has reached 
$m_p=1\;M_J$, but we now let the planet evolve under 
the influence of the disk for $\sim 500$ orbits.\\
The time evolution of Jupiter's semi-major axis is 
shown in the upper panel of Fig. 
\ref{jupiter}. As expected, the planet migrates on a 
timescale corresponding to Type II 
migration in the case where $\hdz/H=0$. For models 
with $\hdz\ne 0$ however, we see here that, 
compared with the Saturn case, the dependency of the 
migration rate upon the value for $\hdz$ 
is stronger, with a clear tendency for the semi-major axis 
to decrease more slowly as the size 
of the dead-zone increases. Examination of the total disk torques, 
which are represented in the lower 
panel of Fig. \ref{jupiter}, clearly reveals that these decrease in 
magnitude with increasing the 
size of the dead-zone. The outer and inner torques exerted 
on the planet are displayed in the 
upper and lower panels of Fig. \ref{torques-jupiter}, respectively. 
Interestingly, and despite a 
weak dependency on the value for $\hdz$, the 
presence of a dead-zone reduces the effects of 
the outer torques exerted on the planet. 
As illustrated in Fig. \ref{surface-density}, which 
shows the disk surface density as a function of radius (at an azimuthal position 
corresponding to that of the planet), this arises because  
the surface density at the outer edge of the gap is greater in the case where 
$\hdz=0$ compared to cases where $\hdz>0$, increasing the 
magnitude of the outer torques. Moreover, Fig.
\ref{surface-density} reveals that the density in the inner 
disk is much more higher for the 
simulation with $\hdz=2.3H$, which consequently favours 
the higher positive torque exerted on 
the planet in this case (see the lower panel of Fig. \ref{torques-jupiter}). 
It should be 
noted that for models with $\hdz\ne 0$, and provided that the 
viscosity is high enough in the 
live-zone, a migration rate which is lower than the 
inward drift velocity in the active region can make 
the disk material from this region flow across the gap. 
This not only enables the inner disk to 
be continuously supplied with gas but also creates a 
positive corotation torque on the planet 
as the gas flows through the gap 
(Masset \& Snellgrove 2001; Morbidelli \& Crida 2007).

\subsection{Planet evolution in an alternative dead-zone model}
\begin{figure}
\centering
\includegraphics[width=0.95\columnwidth]{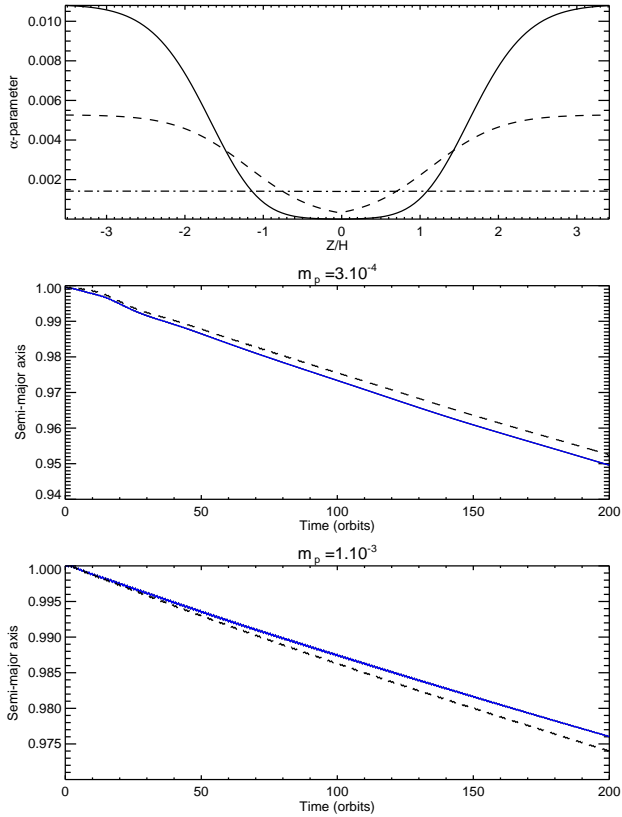}
\caption{{\it Upper panel:} this figure displays the $\alpha$-parameter azimutally-averaged 
vertical profile at $R=1$ for a simulation in which an initial $30\;M_\oplus$ protoplanet grows 
to become a Jupiter-mass planet and in which $\alpha$ is given by Eq. \ref{newalpha}. Different 
times are considered corresponding to times at which $m_p=9.10^{-5}$ (solid line), 
$m_p=3\times 10^{-4}$ (dashed line) and $m_p=10^{-3}$ (dash-dotted line). {\it Middle panel:} this 
figure shows the time evolution of Saturn's semi-major axis, for a model in which $\alpha$ is 
given by Eq. \ref{oldalpha} with $\hdz/H=1.5$ (blue line) and for a simulation in which $\alpha$ is
 given by Eq. \ref{newalpha} (dashed line).
 {\it Lower panel:} same but for Jupiter.}
\label{newrun}  
\end{figure}

The prescription that we have used to model the variation of
viscous stress in disks with dead-zones depends only
on the height of the fluid from the disk midplane.
It is possible that some of our results may be affected
by this prescription, since the spatial extent of a dead-zone should
probably depend on the local column density (assuming that external sources of
ionisation such as cosmic rays or X-rays are responsible 
for ionising the disk). The assumption of a stationary profile
for $\alpha$ is reasonable in a disk which remains largely
unperturbed, but may become inaccurate when gaps form due
to the presence of planets, since material near the midplane
located in the gap should become magnetically active.
To further investigate
this issue, we have performed additional simulations using the 
following vertical profile for $\alpha$:

\begin{equation}
\alpha=\alpha_0(R)\exp\left(-\frac{\eta(z)}{\eta_c}\right)
\label{newalpha}
\end{equation}

\noindent
where $\eta(z)=\int_z^{\infty}\rho dz'$ is the column density at 
height $z$, and where $\alpha_O$
is determined in such a way that the vertically averaged mass flow 
through the disk is 
$\dot M_0$ at each radius. In the previous equation, we set
$\eta_c=8.35\times 10^{-3}$ which corresponds to the initial 
value for the column density at
altitude $z=1.5 H$. The upper panel of Fig. \ref{newrun} 
shows the azimutally-averaged vertical
profile of the $\alpha$-parameter at $R=1$ for a simulation in 
which an initial $30\;M_\oplus$
protoplanet grows to become a Jupiter mass planet, 
and for which $\alpha$ is given by the
previous equation. The solid line depicts the initial profile 
for $\alpha$, whereas the dashed
and dot-dashed lines correspond to times at which 
$m_p=3\times 10^{-4}$ and $m_p=10^{-3}$,
respectively. As the planet grows and is
able to open a gap in the disk, the vertical
profile of $\alpha$ at the position of the planet 
relaxes toward a uniform distribution
with $\alpha=1.4\times 10^{-3}$ throughout the vertical 
extent of the disk. Although not
presented here, it is worthwhile noting that for such a 
simulation, the time evolution of
the planet mass is almost indistinguishable from that 
obtained using the prescription of Eq.
\ref{oldalpha}, as the value for the mass accretion rate is the same.\\

For this model, the middle and lower panels of 
Fig. \ref{newrun}, respectively, display the
evolution of the semi-major axes for the Saturn and Jupiter-mass
planets. Compared with the previous model with 
$\hdz=1.5 H$, the migration of Saturn proceeds very similarly. 
This arises because in both
cases, the value for the laminar viscosity in the active 
region is not high enough for the
corotation torque to be unsaturated. In the Jupiter case,  
the difference between the two
models after $200$ orbits is again marginal despite 
the fact that for the new model $\alpha$ is
vertically constant at the position of Jupiter 
(see the upper panel of Fig. \ref{newrun}). This
supports the idea that the prescription we adopt for 
$\alpha$ has little influence on the
subsequent evolution of the planet.\\

\section{Discussion and conclusion}

In this paper we have presented the results of 
hydrodynamic simulations aimed at studying the 
effect of a dead-zone on the formation and evolution of 
giant planets. For layered 
disks we model the dead-zone as a region where the viscosity 
profile depends on the distance from the equatorial 
plane, and we considered different values for the dead-zone size, 
ranging from  $\hdz=0$ to 
$\hdz=2.3H$. For each value of $\hdz$, we focused on 
a model in which a $30\;M_\oplus$ 
protoplanet accretes gas from the disk until its mass reaches 
either $m_p=1\;M_S$ or $m_p=1\;M_J$.
Once the final planet 
mass has been attained, we assumed that accretion stops, and let the planet 
evolve under the action of disk torques. \\
The results of our simulations indicate that, for a given mass accretion rate 
through the disk, the timescale over which a $30\; M_\oplus$ 
protoplanet grows to become a 
giant planet does not depend on the size of the 
dead-zone. This occurs because 
the planet is able to pull some of the gas 
from the live-zone down toward the midplane,
from where it can be accreted, as 
the latter flows into the gap region. However, we find that the migration of 
giant planets can be modestly affected by the presence of 
a dead-zone. Indeed, there is a clear 
tendency for Jupiter-mass planets to migrate more slowly 
as the size of the dead-zone 
increases. For Saturn-mass planets, we find that 
although migration depends only weakly on 
the size of the dead-zone for models with $\hdz\le 1.5 H$, 
it proceeds more slowly in the case 
where $\hdz=2.3H$. This is due to the fact that in the latter case, 
the viscosity in the 
live-zone is high enough for the corotation torque to become unsaturated. \\

Although the present study suggests that a dead-zone does not
have a significant impact on the growth and 
migration of giant planets, it will be of interest to examine
this issue using three-dimensional 
MHD simulations, which include a dead-zone, in order to 
confirm (or refute) our findings. Of particular interest
will be the issue of what happens when gap opening allows
disk material in the vicinity of the planet to become
magnetically active.
We will address this issue in a future paper.

\end{document}